\documentstyle[11pt,newpasp,twoside]{article}
\def\edcomment#1{\iffalse\marginpar{\raggedright\sl#1\/}\else\relax\fi}
\reversemarginpar

\begin{document}
\title{On the origin of the color Tully-Fisher and 
color-magnitude relations of disk galaxies}

\author{Vladimir Avila-Reese and Claudio Firmani\altaffilmark{1}}
\affil{Instituto de Astronom\'\i a-UNAM, A.P. 70-264, 04510 
M\'exico, D. F.}

\altaffiltext{1}{Also Osservatorio Astronomico di Brera, via E.Bianchi 
46, I-23807 Merate, Italy}

\vspace*{0.7cm}

$\bullet$ The increment of the slope of the T-F relation (TFR)
with the passband wavelength has been called the 
{\bf color TFR}. For example,
from the observed $K-$ and $B-$band TFRs one obtains that  
$L_K/L_B\propto V_m^{0.6}$ (Pierce \& Tully1999, \apj, 387,47).
Since $L_K$ traces the stellar disk mass $M_s$, then $M_s/L_B\propto
V_m^{0.6}$. 
The empirical fact that more luminous galaxies tend to be redder
have been called the {\bf color-magnitude relation}: $(B-H)\propto
\gamma$log$L_B$, $\gamma\approx 0.4-1.2$. 

$\bullet$ {\bf Why the mass-to-luminosity ratio and the color 
index of disk galaxies do depend on $V_m$ (or luminosity)?}
At least there are three alternatives: the star formation (SF) efficiency,
the gas infall efficiency and/or the internal face-on dust extinction 
depend on the galaxy mass. We explore the last
alternative since self-consistent models of disk galaxy
evolution within the hierarchical formation scenario show that the 
$M_s/L_B$ ratio and $B-H$ do not significantly depend on mass or 
luminosity (Avila-Reese \& Firmani 2000, RevMexA\&A, v. 36, in press;
Avila-Reese et al., this volume).

$\bullet$ Observations indeed show that dust and metallicity increase
with $L_B$. In a more quantitative fashion, Wang \& Heckman 
(1996: \apj, 457, 645, WH) have established that the UV(young massive
stars)-to-FIR (the same young stars {\bf +dust absorption}) ratio 
decreases rapidly with $L_B$ $\Rightarrow$ {\bf the dust opacity increases
with $L_B$:} $\tau_B=\tau_{B,o}(L_B/L_{B,o})^\beta$ (eq. 1), with $L_{B,o}=
1.3\times 10^{10}L_{B\odot}$, $\tau_{B,o}=0.8\pm 0.3$, and $\beta=
0.5\pm 0.2$ (WH).  

$\bullet$ Applying the uniform slab model, and using eq. (1) with 
the central values for the constants, a good approximation for the
extinction in the range of $10^8-10^{11}L_{B\odot}$ is: 
$A_B\approx 0.38+0.42$log$L_{B,10}+0.14($log$L_{B,10})^2$ (eq. 2). 
Assuming that in the $H$ band dust absorption is negligible, this result
shows that $(B-H)$ will redden due to dust extinction roughly 
as $\propto 0.42$log$L_B$, which reasonable agrees 
with that is observed. On the other hand, in the understanding that
the origin of the TFR in the different bands is common, the fact that
$L_K/L_B\propto V_m^{0.6}$ is easily accounted for the $B-$band extinction
given by eq. (2). In the hierarchical scenario of galaxy formation
indeed the TFR for any band is a common imprint of the mass-velocity 
relation of the CDM halos.

$\bullet$ In conclusion, {\it the luminosity-dependent dust extinction 
reported by WH easily explains the color TF and color-magnitude
relations of disk galaxies;} there is not necessity to evoke
mass dependent SF and gas infall efficiencies. The extinction 
could depend on mass
because the efficiency of metal ejection out of the disk might
be larger for smaller galaxies and/or because more massive
galaxies have higher surface densities than the smaller ones.

\end{document}